\documentstyle[11pt, epsf]{article}
\newcommand{\vsigman}     {\mbox{\boldmath$\sigma _n$}}
\newcommand{\vi}          {\mbox{\boldmath$I$}}
\newcommand{\vhatkn}      {\mbox{\boldmath$\hat k _n$}}

\newcommand{\sighe}       {\sigma_{^3\!He}}

\newcommand{\cvres}       {resonance}
\newcommand{\lngasym}     {Im $F_3$}
\newcommand{\spnrot}      {Re $F_3$}

\title{Measurement of parity-nonconserving rotation of neutron 
spin in the 0.734-eV p-wave resonance of $^{139}La$}

\author{ T. Haseyama$^a$\footnote{Corresponding author. 
E-mail: haseyama@carrack.kuicr.kyoto-u.ac.jp, 
Fax: +81-774-38-3286, 
Present address: Nuclear Science Research Facility,
Institute for Chemical Research, Kyoto University, 
Gokasho Uji, Kyoto 611-0011, Japan }, 
K. Asahi$^b$,  J.D. Bowman$^c$, P.P.J. Delheij$^d$,
H. Funahashi$^a$, S. Ishimoto$^e$,  G. Jones$^f$\footnote{Present address: 
Hamilton College, Clinton, New York 13323, USA}, \\
A. Masaike$^a$\footnote{Present address: Fukui University of Technology, 
Gakuen 3-6-1 Fukui-shi, Fukui 910-8505, Japan}, 
Y. Masuda$^e$,  Y. Matsuda$^a$\footnote{Present address: Institute 
of Physical and Chemical Research (RIKEN), Hirosawa 2-1 Wako-shi, 
Saitama 351-01, Japan}, 
K. Morimoto$^e$, S. Muto$^e$, S.I. Penttil\"a$^c$,  \\
V.R. Pomeroy$^g$, K. Sakai$^b$, E.I. Sharapov$^h$,
D.A. Smith$^c$\footnote{Present address: 
Stanford Linear Accelerator Center, Stanford, California 94309, USA}\\
and  V.W. Yuan$^c$ \\
$^a$ {\small \it Department of Physics, Kyoto University, 
Kitashirakawa Oiwakecho, Sakyo-ku, Kyoto 606-8502, Japan} \\
$^b$ {\small \it Department of Physics, Tokyo Institute of 
Technology, Oh-okayama 2-12-1, Meguroku, Tokyo, 152-8551, Japan} \\
$^c$ {\small \it Los Alamos National Laboratory, Los Alamos, 
New Mexico 87545, USA} \\
$^d$ {\small \it TRIUMF, Vancouver, British Columbia, 
Canada V6T 2A3} \\
$^e$ {\small \it High Energy Accelerator Research 
Organization (KEK), Oho 1-1, Tsukuba, Ibaraki 305-0801, Japan } \\
$^f$ {\small \it National Institute of Standards and Technology, 
Gaithersburg, MD 20899-3460, USA } \\
$^g$ {\small \it Physics Department,
University of New Hampshire, Durham, NH 03824, USA  } \\
$^h$ {\small \it Joint Institute for Nuclear Research, 
141980 Dubna, Moscow, Region, Russia} \\
}

\begin{document}

\setlength{\baselineskip}{25pt}

\maketitle

\begin{abstract}

The parity nonconserving spin rotation 
of neutrons in the 0.734-eV p-wave resonance of $^{139}La$ 
was measured with the neutron transmission method.  
Two optically polarized $^3He$ cells were used before and behind a
a 5-cm long $^{139}La$ target as a polarizer and an analyzer of 
neutron spin. 
The rotation angle was carefully measured by flipping 
the direction of $^3He$ polarization in the polarizer in sequence. 
The peak-to-peak value of the spin rotation was found to be 
$ ( 7.4 \pm 1.1 ) \times 10^{-3} $ rad/cm 
which was consistent with the previous experiments. 
But the result was statisticallly improved. 
The s-p mixing model gives the weak matrix element as  
$xW = ( 1.71 \pm 0.25 )$ meV.  
The value agrees well with the one deduced from the 
parity-nonconserving longitudinal asymmetry  
in the same resonance. 

\end{abstract}


There have been a number of experimental studies concerning the 
effects of the weak interaction in hadronic systems.  
The ratio of the strength of the weak interaction to 
the strong interaction in such systems is estimated 
to be $G_F m_{\pi}^2 / 4 \pi \sim 2 \times 10^{-7}$, 
where $G_F$ and $m_{\pi}$ are the Fermi coupling constant and 
the mass of pion, respectively.  Observed longitudinal asymmetries of 
the cross sections for proton-proton scattering \cite{balzer,berdoz,yuan} 
are on the order of this estimate.
However, it was found that parity-nonconserving effects 
are tremendously enhanced in p-wave resonances 
of neutron-nucleus systems.  In particular, 
the helicity asymmetry of the cross section at the 0.734-eV 
p-wave resonance of $^{139}La$ was found to reach 10\% 
\cite{masuda,vinnie,shimizu,serebrov}. 
As a theoretical explanation for such large 
parity-nonconserving (PNC) longitudinal 
asymmetries, a model based on the interference between the p-wave 
and its neighboring s-wave resonances was proposed \cite{sushkov}.  

The PNC effect appears in several measurable 
quantities in the neutron-nucleus system
in addition to the longitudinal asymmetry 
of the total cross section of neutron-nucleus interaction. 
One of them is the rotation of neutron spin 
around the momentum vector during transmission in a target material 
while being through matter. 
According to the neutron optics, 
the forward scattering amplitude $f(0)$ 
for the neutron-nucleus scattering is expressed as 
\begin{equation}
f(0) = F_0 + F_1 \vsigman \cdot \vi 
+  F_2  \vsigman \cdot (\vhatkn \times \vi) 
+ F_3 \vsigman \cdot \vhatkn \,\, ,
\end{equation}
where $\vsigman$ denotes the Pauli spin matrices for the neutron, and
$\vi$ and $\vhatkn$ are unit vectors representing the directions of 
the target nuclear spin and the incident neutron momentum, respectively.  
The complex coefficients $F_i (i=0 - 3)$ are functions 
of the neutron energy. 
The terms $ \vsigman \cdot ( \vi \times \vhatkn )$ 
and $ \vsigman \cdot  \vhatkn $
represent the parity nonconserving amplitudes. 
In the case of an unpolarized target, the terms with
$F_1$ and $F_2$ vanish, and 
the forward scattering amplitude is simply described as
\begin{equation}
f(0) = F_0 + F_3 \vsigman \cdot \vhatkn \,\, .
\end{equation}
The spin rotation angle around the momentum vector and the 
difference of the cross sections for the different neutron 
helicities are given as 
\begin{equation}
\frac{\partial \phi}{\partial z} = - \frac{4 \pi}{k_n} N \mbox{Re} F_3\, ,
\label{real F3}
\end{equation}
and
\begin{equation}
\Delta \sigma = \frac{8\pi}{k_n} \mbox{Im} F_3\, ,
\label{imag F3}
\end{equation}
respectively, where $N$ is the number density of the target nuclei  
and $k_n$ is the wave number of the neutron.  
Accordingly, the longitudinal asymmetry and the 
spin-rotation angle are connected with each 
other through the forward scattering amplitude, $F_3$. 

Now we examine the PNC effect on the forward scattering amplitude 
considering the case when an s-wave and a p-wave resonances contribute 
to a neutron transmission for which the measurement is made. 
Due to the presence of a weak matrix element $W$ which connects 
the s- and p-states so that an incident neutron may be captured 
by the target nucleus in s-wave and re-emitted in p-wave,
or vice versa.  In this context, the interference between these two 
processes gives rise to the parity nonconserving amplitude $F_3$. 
Although the total angular momentum $j$ of the p-wave neutron can
have values 1/2 and 3/2, only the $j=1/2$ component contributes 
to the interference because of the conservation of the total angular momentum. 
Thus the s-p mixing model gives $F_3$ as 
\begin{equation}
F_3 = - \frac{g}{k_n} 
        \frac{xW(\Gamma^n_s \Gamma^n_p)^\frac{1}{2}}
        {(E_n-E_s+\frac{i}{2}\Gamma_s)(E_n-E_p+\frac{i}{2}\Gamma_p)} \,\, ,
\label{F3}
\end{equation}
where $E_n$ is the incident neutron energy and $E_s$ ($E_p$), $\Gamma_s$
($\Gamma_p$), and $\Gamma^n_s$ ($\Gamma^n_p$) are the energy, 
total width, and neutron width of the s-wave resonance (p-wave resonance).
\cite{gudkov}  The spin statistical factor $g$ 
is calculated with the spin of the target nucleus $I$ and 
that of the resonance state $J$ as 
\begin{equation}
g = \frac{2J+1}{2(2I+1)} \,\, .
\end{equation}
The factor $x$ is defined as 
\begin{equation}
x^2 = \frac{\Gamma^n_{p_{1/2}}}{\Gamma^n_p} \,\, ,
\end{equation}
where $\Gamma^n_{p_{1/2}}$ is the partial neutron width for 
the p-wave neutrons whose orbital and spin angular momenta make 
a total angular momentum $j=1/2$.
From Eq.(\ref{real F3}) and Eq.(\ref{F3}) and assuming 
$|E_p - E_s| >> \Gamma_s$ and  $\Gamma_s \sim \Gamma_p$, the 
PNC neutron-spin rotation is given as 
\begin{equation}
\frac{\partial \phi}{\partial z} \simeq \frac{4\pi g N}{k_n^2} \,
        \frac{xW(\Gamma^n_s\Gamma^n_p)^\frac{1}{2}}{E_p - E_s} \,
        \frac{E_n-E_p}{(E_n-E_p)^2+\frac{1}{4}\Gamma_p^2} \,\, .
\label{rotation angle from amplitude}
\end{equation}

An attempt to measure the neutron spin rotation angle
covering the 0.734-eV p-wave resonance of $^{139}La$ was performed at KEK 
as reported in the reference\cite{sakai}.  They used a superconducting 
neutron polarimeter, which comprised a neutron spin polarizer 
of polarized protons, a neutron spin analyzer of polarized 3He 
nuclei and Meissner sheets. \cite{masudamag}
They found that the energy dependence of the rotation angle has a dispersive 
behavior changing its sign at the center of the p-wave resonance, 
as expected from the s-p mixing model.  
They deduced the value of $xW$ to be $(1.04 \pm 0.40)$ meV. 
The value is consistent with those obtained from the longitudinal 
asymmetries of the cross section \cite{masuda,shimizu} within the errors. 
Serebrov et al. measured the neutron spin rotation angles at several 
energies around 0.734-eV with a $^{139}La$ target at Gatchina 
\cite{serebrov}.
They used a crystal-diffraction method to produce 
a polarized neutron beam and to analyze the neutron polarization 
after the target.  They also obtained 
the result that the spin rotation angle shows a dispersion curve.
\footnote{Recently Serebrov et al. 
improved these measurements with the result 
which is expected to appear soon\cite{serebrov new}. }

We measured the parity-nonconserving rotation of neutron spin 
in the p-wave resonance of $^{139}La$ with 
a high intensity pulsed neutron beam.  The experiment was 
performed at Manuel Lujan Jr. Neutron Scattering Center in 
Los Alamos Neutron Science Center.  The beamline and experimental 
setup are schematically shown in Fig. \ref{beamline setup}.  
We measured the neutron spin-rotation asymmetry in
transmission through a lanthanum target using two optically polarized
$^3He$ neutron spin filters. One was used as a neutron beam polarizer
and the other to analyze transmitted neutron beam polarization.

We used a cylindrical lanthanum target of 5 cm in length 
along the beam direction and 3 cm in diameter.   
The target was suspended by a rod inside a niobium cylinder 
which was placed in the neutron-beam path.  The niobium
cylinder was cooled with liquid helium to the superconducting state
so that it completely excluded the magnetic field in the cylinder.
The magnetic field seen by the neutron changed nonadiabatically 
at the boundary of the cylinder and the neutron spin did not 
experience any rotation when entering or exiting the cylinder. 
The method and the device were 
developed and used in the experiment at KEK\cite{sakai,masudamag}. 

The polarizer and the analyzer of the neutron spins 
were placed before and behind of the lanthanum target, respectively.  
Both spin filters consisted of $^3He$ gas of 6 atm 
enclosed in cylindrical glass cells.  
A small amount of rubidium was introduced in the cells and 
optically pumped by a circularly polarized light of the wavelength 
of 795 nm in a magnetic field of about 20 gauss. 
We used frequency stabilized diode laser arrays. 
The polarization of $Rb$ atoms were transfered to the $^3He$ nuclear
spins by hyperfine interactions upon atomic collisions.  
Since the neutron absorption in $^3He$ is essentially 
associated with the $0^+$ compound state at low neutron
energies, $^3He$ absorbs only neutrons with the spins antiparallel to
the $^3He$ spins and let the neutrons with parallel spin go through. 
The polarization axes of the two spin filters were set transverse 
to the beam direction and perpendicular to each other. 
Using the adiabatic fast passage method of NMR,  
the $^3He$ polarization of the polarizer was 
flipped every 7 minutes to minimize the systematic errors.
Flipping was carried out with an eight-step sequence of $+ - - + - + + -$, 
where $+$ and $-$ denote the direction of the $^3He$ polarization in the 
polarizer rotated around the beamline by $+90 ^\circ$ or $-90 ^\circ$
with respect to the direction of the polarization of $^3He$ of the analyzer.
The neutron transmission was measured with 4800 beam bursts 
for each flipped state.

The values of absolute polarizations of the $^3He$ filters were obtained 
from the transmission enhancement through the polarized $^3He$ gas.
$^3He$ polarizations of the polarizer and analyzer were
monitored by measuring amplitudes of NMR signals of $^3He$. 
The direction of the polarization of the analyzer was 
fixed during the measurement. 
The ratio of the neutron transmission through polarized $^3He$ gas 
is enhanced over the transmission through unpolarized $^3He$ gas.
The enhancement is given by
$ \cosh \left(P_{^3He}  \sighe N l \right) $, where 
$P_{^3He}$, $\sighe$,  $N$, and $l$ are the polarization, 
the neutron cross section, the number density of the $^3He$ nuclei 
and the thickness of the cell, respectively.  
The neutron transmission enhancement for each 
spin filter was measured without the lanthanum target. 
Typical polarizations of $^3He$ in the polarizer 
and the analyzer were 56\% and 29\%, respectively. 

The neutron spin was kept in the same direction as a spin-holding 
magnet between the polarizer and the target, and also as another 
magnet between the analyzer and the target.

The neutron beam was defined by brass 
collimators with apertures of 23 mm in diameter. 
The collimators were placed upstream of the polarizer 
and downstream of the analyzer. Transmitted neutrons were detected with $^{10}B$ loaded liquid 
scintillation counters located 56-m downstream of the spallation target. 
The liquid scintillator was enclosed in a flat cylinder with
diameter of 43 cm and thickness of 4 cm. 
Scintillation photons were detected with 55 photomultiplier 
tubes of 2 inches in diameter (Amprex XP2262B).  
The detection efficiency for 0.734-eV 
neutrons was nearly 100\%.\cite{yen}  
The signals were counted by multi-channel 
scalers (EG\&G ORTEC Turbo MCS) after passing through discriminators. 
The energies of neutrons were determined by 
time-of-flight (TOF) measurements. The start signal for the 
TOF measurements was given by
the proton pulse incident to the spallation target. The resolution of
the neutron energy at 0.734 eV was about 5 meV.

The neutron detector was sensitive also to $\gamma$-rays and 
neutrons which were scattered with collimators and walls.
They showed wrong TOF timing.  
We studied the background in the detected neutrons 
by inserting foil absorbers of tantalum, indium,
and cadmium in the neutron beam downstream of the analyzer.
The thickness of the foils were selected so that neutrons in the
resonances of 0.18 eV for $^{113}Cd$, 1.46 eV for  $^{115}In$ 
and  4.3 eV for $^{181}Ta$ were completely absorbed, 
so these being black resonances in the measurement, 
thus detector counts at these resonance energies came entirely from
the background.  The background levels at three black 
resonances were estimated by interpolating the 
count rates at the three black resonances.

The neutron flux was monitored with a pair of 
ion chambers placed upstream of the polarizer.  
One of them was filled with $^3He$ gas, while the other with $^4He$.  
Both of the chambers were operated at the pressure of 1 atm. 
Each chamber had a charge collection plate at the center, which 
was sandwiched between two anode plates hold at about +600 V.  
The thickness of the sensitive area between the two anodes was 2 cm. 
Difference of the currents of the two chambers gives the
neutron flux without $\gamma$-ray contribution, since the $^3He$ 
chamber is sensitive to both neutrons and $\gamma$-rays, whereas 
the $^4He$ chamber is sensitive only to $\gamma$-rays.

The ''flipping ratio'' is defined as the asymmetry of the neutron 
transmissions for two directions of the polarizer. 
It is written as the product of the rotation angle of the neutron spin 
and the total analyzing power determined by $^3He$ polarization. 
The latter is given as 
$\tanh  \left(P_{pol} \sighe N_{pol} l_{pol} \right)
\tanh \left(P_{ana} \sighe N_{ana} l_{ana} \right)$,
where $P_{pol}$, $N_{pol}$ and $l_{pol}$ are the polarization, 
neutron cross section, number density and  
length of the $^3He$ polarizer, respectively 
and $P_{ana}$, $N_{ana}$ and $l_{ana}$ are those of the analyzer. 
Actually we determined the factors given by the hyperbolic 
functions experimentally from the neutron transmission enhancement 
and NMR signals.

\begin{figure}[htb]
\epsfile{file=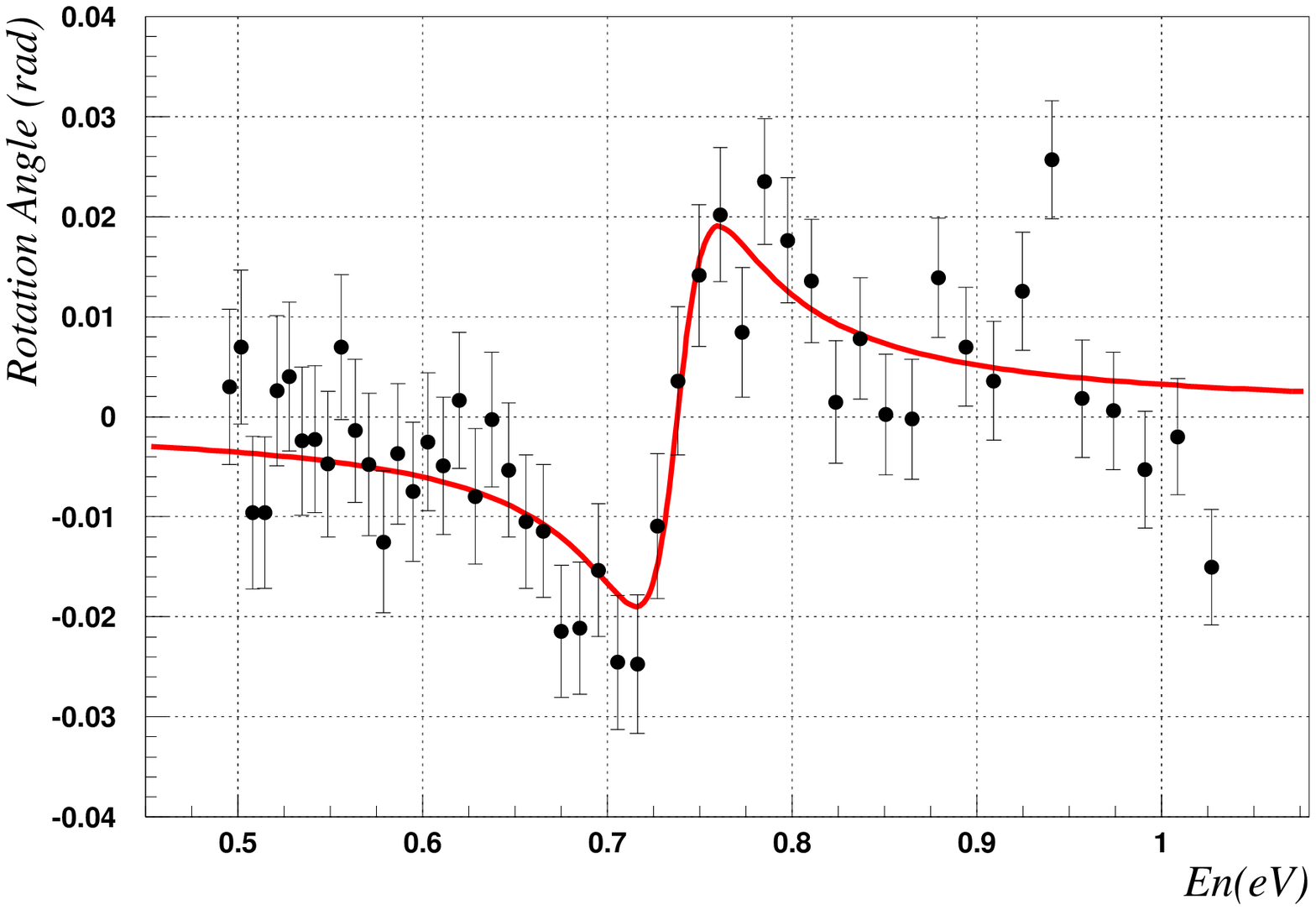,scale=0.6}
\caption{
The neutron-spin rotation angle in the 0.734-eV p-wave resonance 
of $^{139}La$.  Statistical errors of the rotation angle are 
shown as bars.  The solid line represents the best fit of the 
first term of Eq.(\ref{angle extraction function}) to the 
experimental values which were obtained by averaging the data taken 
in the two different magnetic fields. 
The difference between the maximum and the minimum of 
the solid line is $ ( 3.72 \pm 0.55 ) \times 10^{-2} $ rad
after subtracting off $S \Phi_{bg}$.
} 
\label{dispersion curve}
\end{figure}

The spin rotation angle $\Phi$ measured with the lanthanum target was 
fitted for a function 
\begin{equation}
	\Phi = 
        \phi\cdot 
        \frac{\frac{E_n - E_p}{\Gamma_p/2}}
        {1+ \left( \frac{E_n - E_p}{\Gamma_p/2} \right)^2} 
        + C + S \Phi_{bg} \,\, ,
\label{angle extraction function}
\end{equation}
where $\phi$, $C$, $S$ are fitting parameters.
The first term represents the energy dependence of the neutron spin 
rotation.  The second term, $C$ is the constant which indicates the 
error in the fine adjustment of the direction of polarization of 
the polarizer and analyzer.  
The third term, $\Phi_{bg}$ represents a misalignment effect, 
which originates from field inhomogeneity.  A transverse component 
of the neutron polarization to the spin holding field appears upon 
passage through the region where field direction rapidly varies.
We measured $\Phi_{bg}$ without the lanthanum target.  However, 
the experimental condition might be changed when we placed a 
thick target. For example, slight deflections of the neutron beam path 
due to multiple small-angle scatterings, or a failure in $^3He$ 
polarization monitor probably caused a small change in $\Phi_{bg}$.  
We assumed that $S$, whose deviation from unity represents the change 
in $\Phi_{bg}$, is independent of the neutron time of flight.  
According to Eq.(\ref{rotation angle from amplitude}) 
$\phi$ is related to the weak matrix element $W$ by the equation 
\begin{equation}
\phi = \frac{8\pi g N l}{k_n^2 \Gamma_p} \,
        \frac{xW(\Gamma^n_s\Gamma^n_p)^\frac{1}{2}}{E_p - E_s} \,\, .
\end{equation}
The misalignment effect, $\Phi_{bg}$ depends on the neutron time 
of flight. The dependence is affected by the holding field 
distribution in the neutron beam path. The fitting was carried out 
for the two sets of data which were obtained with magnetic field of
20 and 15 Gauss at the spin-holding magnets. 
In the higher-field case, the fitting parameters were obtained as 
\begin{equation}
        \phi =  ( 3.99  \pm 0.55 )  \times 10^{-2} \,\, \mbox{rad}, \,\,
        C    =  ( 3.79  \pm 0.95 )  \times 10^{-3} \,\, , \mbox{and} \,\,
        S    =  1.115 \pm 0.012 \,\,.
\end{equation}
In the lower-field case, the results were 
\begin{equation}
        \phi = ( 3.42  \pm 0.76 ) \times 10^{-2} \,\, \mbox{rad} , \,\,
        C    = (-3.26  \pm 1.42 ) \times 10^{-3} \,\, , \mbox{and} \,\,
        S    = 1.041  \pm 0.011 \,\, .
\end{equation}
The errors shown above were obtained statistically by fitting.  
The two values of phi for the two holding field conditions 
were consistent with each other.  We took the averaged value 
of these results for $\phi$.  The neutron spin rotation angle 
in the p-wave resonance of $^{139}La$ target 
is shown as a function of the neutron energy in Fig. \ref{dispersion curve}. 

The difference between the maximum and the minimum values of 
the rotation angle in a target of a unit length was obtained as 
$ ( 7.4 \pm 1.1 ) \times 10^{-3} $ rad/cm.
It should be mentioned that in the present analysis 
the systematic error caused by the $^3He$ polarization measurement 
is not explicitly shown.  However, the possible error in the fitting 
caused by the ambiguity originating from the polarization measurement 
are negligible compared to the error due to the ambiguities of the 
fitting parameters.  The systematic errors originate from the 
uncertainties in the NMR sensitivity and in the angular settings of 
axes for the two spin filters.  

In order to extract $xW$ from the rotation angle,
the energy $E_p$ and the width $\Gamma_p$ of the resonance
were determined from this experiment. 
The width of the s-wave resonance $\Gamma_s$ was 
considered to be negligible with respect to $( E_n - E_s )$ in the 
vicinity of the p-wave resonance.  We used the values 
listed in \cite{mughabghab} for the other parameters.
With these values, we obtained 
\begin{equation}
xW = ( 1.71 \pm 0.25 \, ) \,\, \mbox{meV} \,\, .
\label{final result xW}
\end{equation}

The observed spin rotation angle $\phi$ exhibits a dispersion curve 
whose center and width were independently determined from the 
neutron absorption experiment.  The value of $xW$ agrees well with 
those obtained from the PNC longitudinal asymmetries in the same 
resonance as shown in Table \ref{xW list}.
In the present experiment we have confirmed that the large 
enhancement of PNC effect in the p-wave resonance of 
$^{139}La$ exists in the neutron spin rotation. 
The enhancements in the longitudinal asymmetry 
and the spin rotation can be described consistently 
by the s-p mixing model.

The authors would like to thank the LANSCE staff for
their support during the experiment. 
We acknowledge Prof. H. En'yo for his valuable advice in the data analyses. 
We express our gratitude to the late professor H. Ikeda for 
the preparation of the experimental apparatus.
This work has benefited from the use of LANSCE which is operated 
under Contract W-7405-ENG-36. 
We acknowledge the financial supports of 
Grant-in-Aid for scientific research numbers 09559017 and 09044105.
One of the authors (T.H.) was supported by 
JSPS Research Fellowships for Young Scientists 
and the GRA program of Los Alamos National Laboratory.

\begin{table}
\begin{center}
\begin{tabular}{|c|c|c|c|}
\hline \vspace{-4.5mm} &&& \\
$ xW(\mbox{meV}) $ & observable & 
energy region & reference \\
\hline \vspace{-4.5mm} &&& \\ \hline
 $ 1.7  \pm 0.1   $ & \lngasym & \cvres  &\cite{shimizu} \\
 $ 1.68 \pm 0.06  $ & \lngasym & \cvres  &\cite{vinnie}  \\
 $ 1.77 \pm 0.13  $ & \lngasym & \cvres  &\cite{serebrov}\\
\hline 
 $ 1.63 \pm 0.21  $ & \spnrot  & cold    &\cite{heckel}  \\
 $ 1.04 \pm 0.40  $ & \spnrot  & \cvres  &\cite{sakai}   \\
 $ 1.71 \pm 0.25  $ & \spnrot  & \cvres  & present work  \\
\hline
\end{tabular}
\caption{Experimental values of $xW$ for 0.734-eV p-wave resonance of $^{139}La$.}
\label{xW list}
\end{center}
\end{table}

\clearpage
\onecolumn
\center

\begin{figure}
\begin{center}
\epsfile{file=fig1.eps,scale=0.5}
\caption{Experimental setup.}
\label{beamline setup}
\end{center}
\end{figure}


\clearpage

\begin{thebibliography}{l}
\bibitem{balzer}   R. Balzer et al.,        Phys. Rev.      C30,1409(1984)
\bibitem{berdoz}   A.R. Berdoz et al., Nucl. Phys. A629,433(1998)
\bibitem{yuan}     V. Yuan et al., Phys. Rev. Lett 57, 1680(1986)

\bibitem{masuda}   Y. Masuda et al.,        Nucl.Phys.      A478,737(1988)
\bibitem{vinnie}   V.W. Yuan et al.,   Phys.Rev.     C44,2187(1991)
\bibitem{shimizu}  H.M. Shimizu et al.,     Nucl.Phys.      A552,293(1993)
\bibitem{serebrov} A.P. Serebrov et al.,  JETP Lett. 62,545(1995), 

\bibitem{sushkov}  O.P. Sushkov and V.V. Flambaum, JETP Letters 32, 352(1980)

\bibitem{gudkov}  V.P. Gudkov and V.E. Bunakov,  Z. Phys. A303,285(1981)

\bibitem{heckel}   B. Heckel et al., Phys.Rev.C29 , 2389(1984)
\bibitem{sakai}    K. Sakai et al.,         Phys.Let. B391,11(1997)

\bibitem{masudamag} Y.Masuda, Neutron Research 1, 53(1993)
\bibitem{serebrov new} A. Petukhov, private communications. 
\bibitem{yen}      Y.-F. Yen et al.,      NIM A 447, 476(2000)

\bibitem{mughabghab}  S.F. Mughabghab et al., Neutron Cross Sections, vol. 1A an
d 1B, (Academic Press, New York) (1988)
\bibitem{Rich}     D. R. Rich, et al., arXiv:physics/9908050 25 Aug 1999
\end{thebibliography}
\end{document}